\documentclass[12pt,letter]{iopart}
\usepackage{graphicx}

\newcommand{\M}{\ensuremath{\mathcal{M}}}               % Phase space
\newcommand{\E}{\mathcal{E}}                            % Energy surface
\newcommand{\wf}{\ensuremath{\mbox{\large$\psi$}}}      % Wavefunction
\newcommand{\abs}[1]{\left\vert#1\right\vert}
\newcommand{\applss}{\,\lower0.5ex\hbox{$\sim$}\kern-0.79em\raise0.5ex\hbox{$<$}\,}
\newcommand{\appgtr}{\,\lower0.5ex\hbox{$\sim$}\kern-0.79em\raise0.5ex\hbox{$>$}\,}
\renewcommand{\(}{\left(}
\renewcommand{\)}{\right)}

\begin{document}

\title[Magnetic billiards as ratchets]{Directed Transport in classical
  and quantum chaotic billiards}

\author{W Acevedo and T Dittrich}

\address{Departamento de F\'\i sica, Universidad Nacional de Colombia,
and \\ CeiBA -- Complejidad, Bogot\'a D.C., Colombia}
\eads{\mailto{jwacevedov@unal.edu.co},\mailto{tdittrich@unal.edu.co}}
\begin{abstract}
We construct an autonomous chaotic Hamiltonian ratchet as a
channel billiard subdivided by equidistant walls attached
perpendicularly to one side of the channel, leaving an opening on
the opposite side. A static homogeneous magnetic field penetrating
the billiard breaks time-reversal invariance and renders the classical
motion partially chaotic. We show that the classical dynamics exhibits
directed transport, owing to the asymmetric distribution of regular
regions in phase space. The billiard is quantized by a numerical
method based on a finite-element algorithm combined with the Landau
gauge and the Bloch formalism for periodic potentials. We discuss
features of the billiard eigenstates such as node lines and vortices
in the probability flow. Evidence for directed quantum transport,
inherited from the corresponding features of the classical dynamics,
is presented in terms of level-velocity statistics.

\end{abstract}

%Uncomment for PACS numbers title message
% Keywords required only for MST, PB, PMB, PM, JOA, JOB?
%\vspace{2pc}
%\noindent{\it Keywords}: Article preparation, IOP journals
% Uncomment for Submitted to journal title message
%\submitto{\JPA}
% Comment out if separate title page not required
\maketitle

\section{Introduction}
Within two decades, the concept of ratchets has undergone a
remarkable process of refinements and transformations. Originating
in an attempt to understand the functioning of motor molecules in
terms of fundamental principles of statistical mechanics, ratchets
have first been conceived as overdamped stochastic systems,
modelling molecular motion in the intracellular medium
\cite{Mad93,AB94,JAP97,Rei02}. Inertia terms have sub\-se\-quent\-ly
been included in the equations of motion to allow for a more complex
dynamics than mere relaxation \cite{JKH96}. As a more radical step,
friction has then been neglected altogether, giving way to Hamiltonian
ratchets \cite{SO&01,SDK05}. In replacing viscous forces, another
mechanism to break time-reversal invariance (TRI) had to be
introduced. It was found in a periodic external driving with
asymmetric time dependence, for example with a sawtooth profile
similar to that of the static ratchet potential \cite{FYZ01}. In this
way, the external driving attained three functions that are crucial
for the behaviour of Hamiltonian ratchets, namely (i), breaking TRI,
(ii), rendering the dynamics chaotic, and (iii), injecting or
absorbing energy in case that a finite mean potential gradient be
present.

Directed transport in classical as well as quantum Hamiltonian
ratchets has been studied in great detail in recent years
\cite{SO&01,SDK05,FYZ01}, which led to various basic insights such as
a sum rule for transport, as a lemma of the conservation of
phase-space volume, dynamical mechanisms that generate directed
currents, and the fingerprints of these currents in the band spectra
of quantum ratchets.

It is then tempting to ask whether ratchet-like transport can be
obtained even {\it abandoning the time-dependent external force\/},
thus making the last step towards {\it autonomous\/} Hamiltonian
ratchets. In view of the functions of the driving mentioned above, it
is clear that alternative means are then needed to destroy TRI and to
produce a chaotic dynamics. This latter requirement implies in
particular that a minimum number of two degrees of freedom is
necessary. By a fortunate coincidence, both conditions can be
fulfilled simultaneously if a static magnetic field is imposed that
penetrates the plane of motion perpendicularly (assuming that the
particles at hand carry an electric charge).

In this paper we study periodic billiards immersed in a
homogeneous magnetic field. In order to remain as close as
possible to one-dimensional ratchet models, we construct them as
quasi-one-dimensional billiard chains. In fact it is sufficient to
consider a straight channel with equidistant walls or diaphragms
attached orthogonally to one sidewall of the channel, leaving a
sequence of gaps on the opposite side. In fact, this configuration
has already been conceived as a model for quantum chaos
\cite{Smi90}. Periodic billiards without magnetic field, where
classically chaotic motion is induced instead by curved walls,
have been studied to elucidate, in particular, fingerprints of
chaos in the band structure and energy eigenstates \cite{LN&96}.

In presence of the magnetic field, straight walls at right angles are
sufficient to render the motion of charged particles chaotic, at least
in parts of phase space, in a billiard that would otherwise be
integrable \cite{RB85,MBC93,BK96}. Upon quantization, classically
chaotic billiards in a magnetic field exhibit signatures of chaos in
observable phenomena like the magnetic susceptibility \cite{NT88}.
Particularly close in spirit to our work are studies of
magnetotransport in ballistic two-dimensional electron gases moving in
antidot superlattices \cite{SR94}. Features of the classical nonlinear
dynamics are reflected in anomalies both of the longitudinal
magnetoresistance \cite{FGK92} and of the Hall effect \cite{GKS92}
that had previously been observed experimentally \cite{WR&91}. A
remarkable duality between the interior (quantum dot) and the exterior
(antidot) states of electron billiards has been elucidated in
\cite{HS02}.

Here, by contrast, we are interested in directed (dc) transport
{\it without\/} applied external voltage. It is the broken
symmetry in the transverse direction together with the breaking of
TRI by the magnetic field that leads to an effective ratchet
behavior. A trajectory transporting, say, to the right, is then
generally not compensated by a corresponding symmetry-related
trajectory transporting to the left. Net directed currents ensue.
This effect has been studied on the classical level for a billiard
chain very similar to our model, though with a smoothly curved
boundary that would induce chaos even in the absence of a magnetic
field \cite{SP05}. Experimentally, ratchet effects have been
observed in quasi-one-dimensional semiconductor superlattices, but
subject to an external periodic driving \cite{HL&01}.

In driven ratchets, even pumping in the sense of directed transport
``uphill'' is possible, as well as stationary currents ``downhill''
\cite{SDK05}. In our case, the fact that we are dealing with an
autonomous system prevents this option. The lack of an inexhaustible
energy source/sink coupled to the system implies that the ratchet
effect depends much more sensitively on the absence of a mean
potential gradient than in driven systems. To be sure, this is true
only for an infinitely extended billiard. For a section of finite
length the zero-gradient transport properties may well affect the
response to an external voltage and will then be reflected in the
system's magnetotransport properties. This aspect is beyond the scope
of the present paper, but we briefly comment on it in the conclusion.

In Section 2, we define the geometry of our magnetic billiard and
construct the classical dynamics as a discrete map based on a
Poincar\'e surface of section. We de\-mon\-stra\-te that directed
currents occur under the same conditions and depend on the same
mechanisms (regular regions embedded in the chaotic sea that are
not symmetric under momentum inversion) as in driven ratchets. The
asymptotic behavior for weak vs.\ strong magnetic field and for wide
vs.\ narrow openings between the billiard compartments is discussed
and its implications for transport are pointed out. For a large region
in parameter space we can even give an approximate analytic expression
for the net current.

Section 3 is dedicated to the quantization of the magnetic
billiard. In some lim\-its (e.g., square billiard, empty channel,
Landau states in the 2-dim.\ electron gas), analytical solutions are
available. They serve as reference for the calibration of the
numerical approaches we have to resort to in general. We briefly
explain the quantization algorithm of our choice, a finite element
method, and how it has to be complemented with Bloch theory as it
applies to the periodic billiard. We discuss features of general
interest of the billiard's band spectrum, e.g., the traces of Landau
levels and diabatic bands associated to invariant manifolds of the
classical phase space, as well as of the eigenstates, such as edge
states and vortices in the quantum flow. Evidence of directed
currents is presented in terms of level-velocity statistics.

We conclude in Section 4 considering possible extensions of this work
and open questions related to them, like relations to the quantum Hall
effect, the construction of pumps from finite sections of the billiard
chain, and the effects of an external driving.

\section{Classical magnetic billiard}

\subsection{Billiard geometry and trajectories}
\begin{figure}[ht!]
\centering
\includegraphics[width=\textwidth]{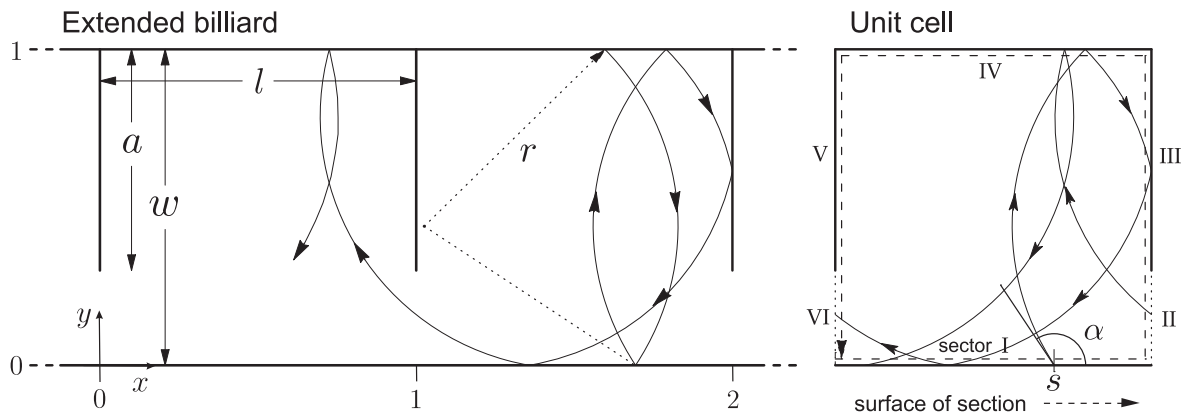}
\caption{Geometry of the billiard.}
\label{fi:shape}
\end{figure}

The billiard is constructed as a straight channel of width $w$
with equidistant barriers of length $a$, separated by a distance
$l$ from each other and attached at right angles to one side
(figure \ref{fi:shape}) \cite{AD03,Ace04}. It is threaded by a
homogeneous magnetic field $B$ perpendicular to the plane of the
billiard. We shall only consider the case of particles with unit
mass and charge ($m=q=1$) moving in a square unit cell ($l=w=1$)
so that barrier size $a$ and cyclotron radius $r_{\rm c} = p/B$,
with $p$, the constant modulus of the momentum, remain as control
parameters. The direction of the magnetic field is chosen such
that for $r_{\rm c} > 0$, clockwise motion on circular-arc-shaped
trajectories results. Upon hitting a wall, trajectories undergo
specular reflection.

\subsection{Symmetries}
The spatio-temporal symmetries of the billiard will attain their full
impact only in the context of quantization, but for completeness we
present them already here.

Considering only the geometry of the walls, the billiard is periodic
in $x$ with period 1,
\begin{equation}\label{eq:perix}
V(x+1,y) = V(x,y),
\end{equation}
and invariant under reflection in $x$,
\begin{equation}\label{eq:reflx}
V(-x,y) = V(x,y).
\end{equation}
Together with (\ref{eq:perix}) this implies reflection symmetry with
respect to every $x_n = n/2$, $n$ integer.

With a non-zero magnetic field, the Hamiltonian retains the
periodicity (\ref{eq:perix}) while parity (\ref{eq:reflx}) is broken,
as is of course time-reversal invariance. Limiting cases of interest
are the empty channel, $a = 0$, where the dynamics is invariant under
continuous translations in $x$, and the closed square billiard which
in the presence of a magnetic field retains a $C_4$ rotation symmetry
(without reflection) \cite{Ham64}.

\subsection{Poincar\'e map, phase-space structure and limiting cases}
A Poincar\'e surface of section is constructed on basis of a closed
line following the entire circumference of the unit cell but shifted
inwards by an infinitesimal amount, see figure \ref{fi:shape}. It is
parameterized by the Birkhoff coordinates $s$ (position along the
cir\-cum\-fer\-ence, measured counterclockwise from the origin) and
$p_s=\cos \alpha$ (tangential momentum after crossing the
surface). The elementary geometry of the trajectories (circular arcs
connected by specular reflection at the walls) allows us to formulate
the dynamics as a Poincar\'e map in terms of a relatively simple
algorithm, see appendix A.

\begin{figure}[ht!]
\centering
\includegraphics[width=0.9\textwidth]{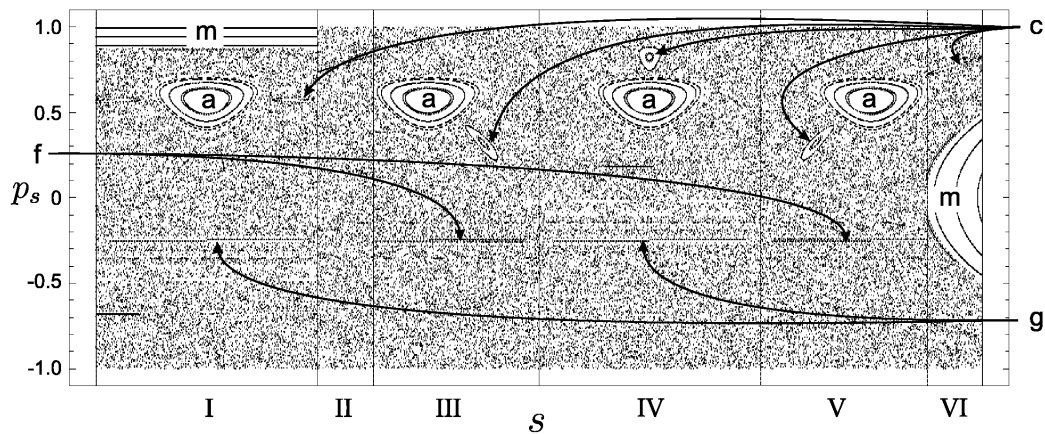}
\caption{Phase portrait for $r_{\rm c}=2$ and $a=0.75$. Lettered
features correspond to the orbit types shown in figure
\protect\ref{fi:po} below.} \label{fi:pss}
\end{figure}

According to the different types of boundary conditions around the
unit cell, we subdivide the Poincar\'e surface of section into six
subsections I to VI (lower wall, right opening, right wall, upper
wall, left wall, left opening). Figure \ref{fi:pss} shows a typical
phase portrait, indicating a mixed dynamics with regular islands
immersed in a chaotic sea and continuous sets of marginally stable
orbits inherited from the integrable motion that dominates in certain
limits, as discussed in the sequel.

As the billiard boundary does not include any curved section, it
is clear that for vanishing magnetic field (divergent radius of
curvature $r_{\rm c}$ of the trajectories), the dynamics cannot be
chaotic. The presence of openings between the cells however render
the phase-space topology of the unit cell multiply connected and
therefore the motion pseudo-integrable in this limit
\cite{RB81,HM90,Wie00}. In this case, in fact, the billiard can be
extended to the entire $(x,y)$-plane by repeated reflection at the
upper and lower horizontal walls, and thus reduces to an infinite
rectangular lattice of straight-wall segments (``barrier
billiard'' \cite{HM90}). This feature is lost for finite $r_{\rm
c}$, as the curvature changes sign upon reflection.

\begin{figure}[ht!]
\includegraphics[width=\textwidth]{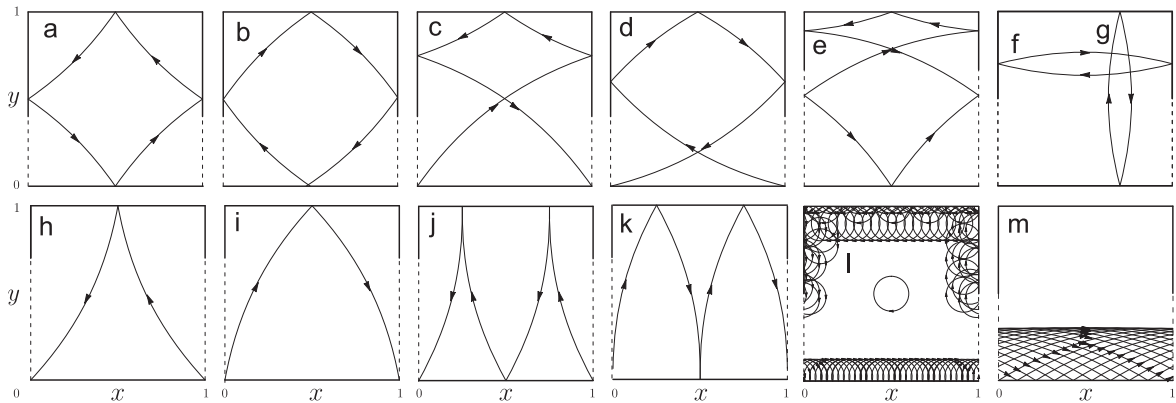}
\caption{Some periodic orbits for $r_{\rm c}=2$ (a -- k), orbits
for $r_{\rm c}=0.1$ (l) and non-contractible trajectories (m). The
location of some of these orbits is indicated by the same letters
in the phase portrait, figure \protect\ref{fi:pss}.} \label{fi:po}
\end{figure}

\begin{figure}[ht!]
\includegraphics[width=\textwidth]{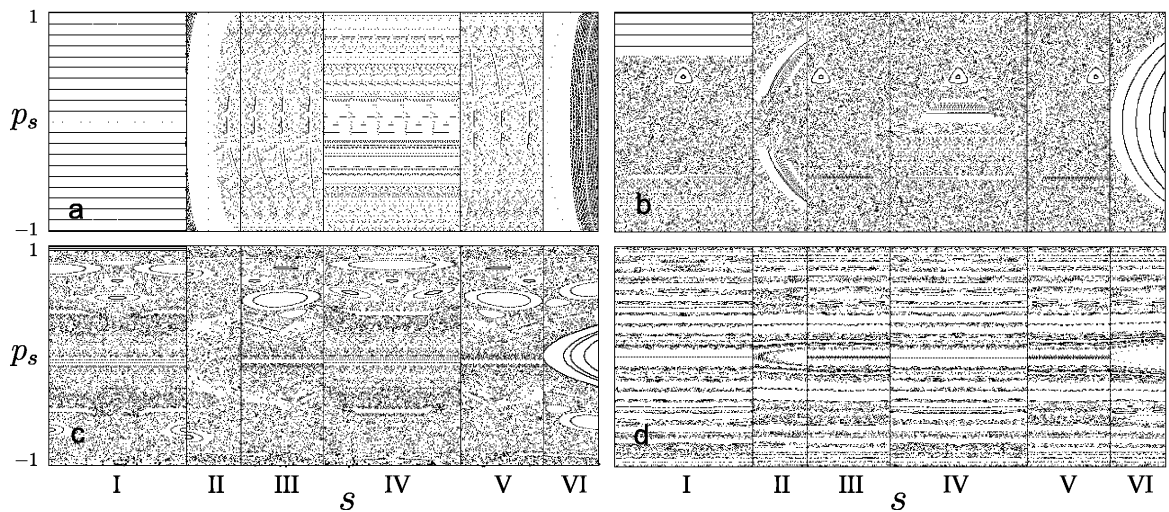}
\caption{Phase-space portraits for $a=0.6$ and $r_{\rm c}=0.1$
(a), $1$ (b), $10$ (c), $100$ (d).} \label{fi:revolv}
\end{figure}

Pseudo-integrability is approached continuously via the
Kolmogorov-Arnol'd-Moser scenario \cite{LL92} for $r_{\rm c} \to
\infty$, see figure \ref{fi:revolv}d. For $r_{\rm c} \to 0$ (the
high-field limit), on the other hand, trajectories touching the
boundary become skipping orbits \cite{HS02,Hal82,Kra98}, while those
inside the billiard reduce to full circles (figure \ref{fi:po}l). In
this limit the system thus approaches integrability, but in a highly
singular manner \cite{Ace04}, cf.\ figure \ref{fi:revolv}a.

\begin{figure}[ht!]
\includegraphics[width=\textwidth]{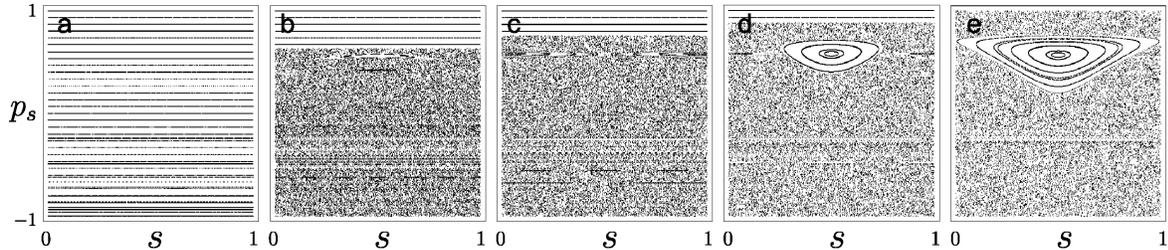}
\caption{Phase-space portraits (sector I only, cf.\ figure
\protect\ref{fi:revolv}) for $r_{\rm c} = 2$ and $a = 0$ (a),
$0.25$ (b), $0.5$ (c), $0.75$ (d), $1$ (e).} \label{fi:aevolv}
\end{figure}

Concerning the wall size $a$, the limit $a = 0$, that is the free
channel billiard, is integrable irrespective of the cyclotron radius.
The closed magnetic square billiard in turn, corresponding to $a =
1$, is integrable in the low- and high-field limits but shows a mixed
dynamics for intermediate values of $r_{\rm c}$ \cite{BK96}. The full
transition from the free channel to the square billiard, for an
intermediate field ($r_{\rm c} = 2$), is illustrated by figure
\ref{fi:aevolv}.

A special feature related to the rectangular geometry of the
billiard is the occurrence of periodic orbits in pairs, one of
them stable, the other unstable, such as figures \ref{fi:po}a vs.\
b and c vs.\ d. They emerge through a tangent bifurcation
\cite{Ott93} from periodic orbits of the integrable motion at zero
field that fulfill certain symmetry conditions, e.g., a
diamond-shaped orbit in the case of figure \ref{fi:po}a/b.
Switching on the field, each straight segment of the orbit can
then be replaced by circular arcs with either curvature, convex or
concave from outside, without affecting the topology of the orbit,
hence their appearance in pairs \cite{Ace04}. Concave orbit
sections tend to stabilize the orbit (panels a, c in figure
\ref{fi:po}), while convex sections tend to destabilize it (panels b,
d). This mechanism is subject to strong pruning at finite values of
the field and may be restricted to extremely small curvature for
longer periodic orbits.

Lifting the condition $l = w$ of square unit cells affects periodic
orbits with $C_4$ symmetry. They do not disappear immediately for $l
\neq w$, but become deformed and eventually are eliminated through
pruning. At the same time, towards the extremes $l \gg w$ or $l \ll
w$, bouncing-ball orbits between the horizontal or vertical walls
(items f,g in figure \ref{fi:po}), respectively, become dominant. 
While they themselves form a set of measure zero (straight lines in
the Poincar\'e section \ref{fi:pss}), they tend to induce a reduction
of the Lyapunov exponent on average of chaotic trajectories in the
surrounding chaotic sea. For the effects on transport of lifting the
square symmetry of the unit cell, see section \ref{se:dirtrans}. 

For intermediate values of $r_{\rm c}$ and $a$, the classical
motion in the billiard is characterized by the coexistence of
regular and chaotic regions in phase space, as reflected in the
Poincar\'e sections, figure \ref{fi:pss}. The regular components
come in two forms: trajectories that result from a mere
deformation of free motion along the open part of the channel, and
therefore represent a continuum of momenta, cf.\ panel m in figure
\ref{fi:po}, and regular islands centered over stable periodic
orbits that can be characterized by rational winding numbers
(panels a, c, e in figure \ref{fi:po}). In topological terms, the
latter are analogous to smoothly contractible orbits in systems
with a cylindrical phase space like the kicked rotor \cite{LL92},
while the former ones cannot be contracted to a point. Both types
of phase-space components contribute in qualitatively different
ways to directed transport, see section \ref{se:dirtrans}.

Based on these distinctions we decompose the energy shell $\E$ into
a complete set of disjunct invariant manifolds $\M_i$:
\begin{equation}
\E=\cup_i \M_i, \quad \M_i \cap \M_j=\emptyset \quad
{\rm for}\quad i\neq j
\label{eq:EnergyShell}
\end{equation}
We select the invariant manifolds \emph{Regular$^0$},
\emph{Regular$^+$}, \emph{Regular$^-$} (regular with zero,
positive, and negative winding number, resp.), and \emph{Chaotic}.
This classification is to a certain degree arbitrary and could
well be finer or cruder. It is, however, appropriate for the
analysis of transport properties in the subsequent section.

\subsection{Directed transport}\label{se:dirtrans}
Directed transport in a billiard with compact phase space is
restricted by a sum rule discussed in \cite{SO&01,SDK05}. It implies
that on average over the total phase space, all directed currents
vanish. In the present case of an autonomous system, the energy shells
form compact invariant subsets of phase space, so that the sum rule
even applies per energy shell $\E$ and reads
\begin{eqnarray}
\tau_\E &=& \sum_i \tau_{\M_i} = 0 \label{eq:sumrule}\\
\tau_{\M_i} &=& \lim_{t\to\infty} \frac{1}{t} \int_0^t {\rm d}t'
\int_{\M_i}{\rm d}^{2f-1}r \rho_t(\mathbf{r}) v_x,
\label{eq:transEnergyShell}
\end{eqnarray}
where $\tau_\E$ is the total transport over the energy shell,
$\tau_{\M_i}$ the transport per invariant set $\M_i$, and
$\rho(\mathbf{r})$ is the natural measure at some point $\mathbf{r}$
on the $(2f-1)$-dimensional energy shell. It corresponds to a
homogeneous distribution over the spatial unit cell and an isotropic
distribution of velocities.

\begin{figure}[h!]
\includegraphics[width=0.49\textwidth]{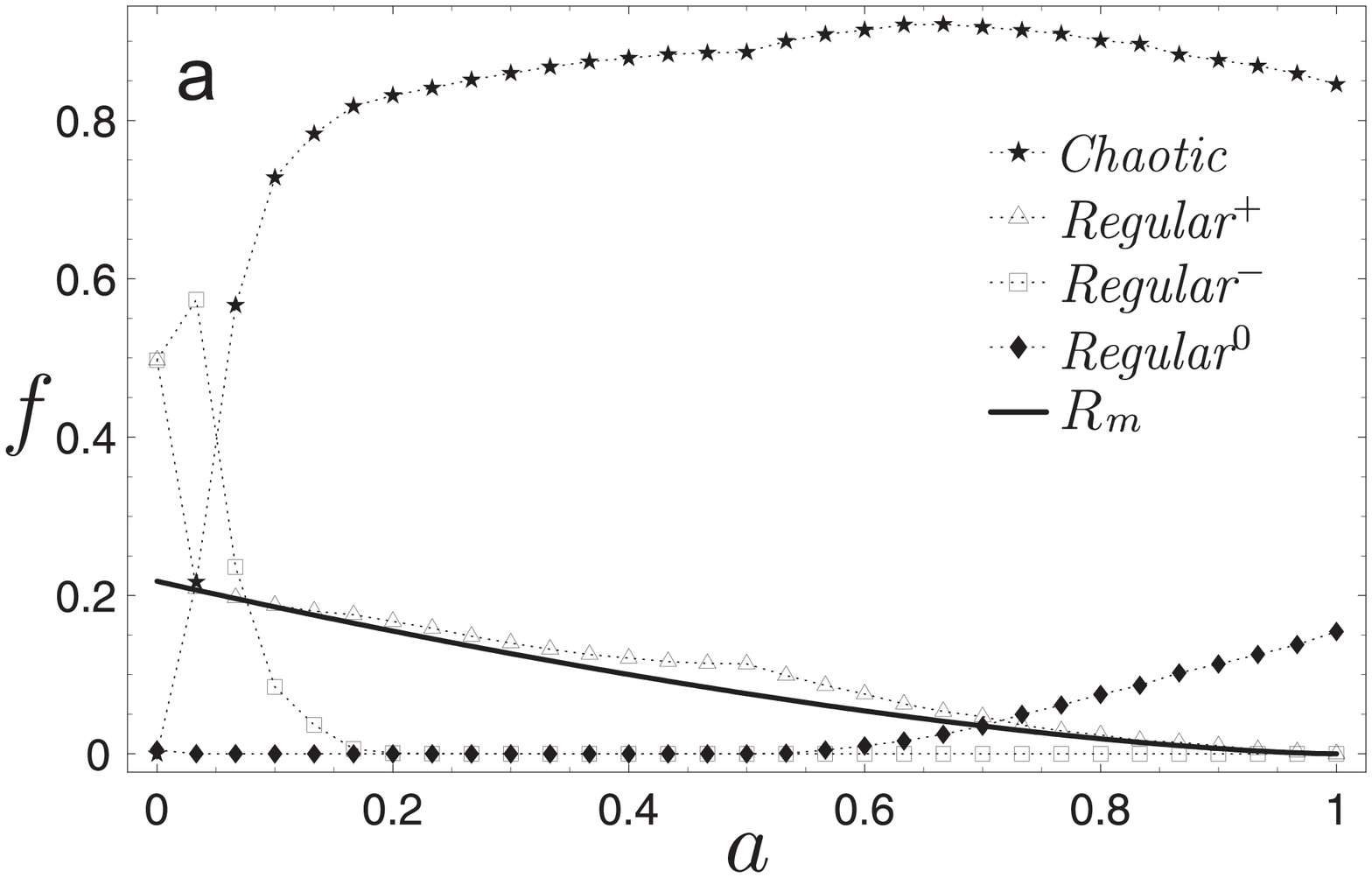}
\includegraphics[width=0.49\textwidth]{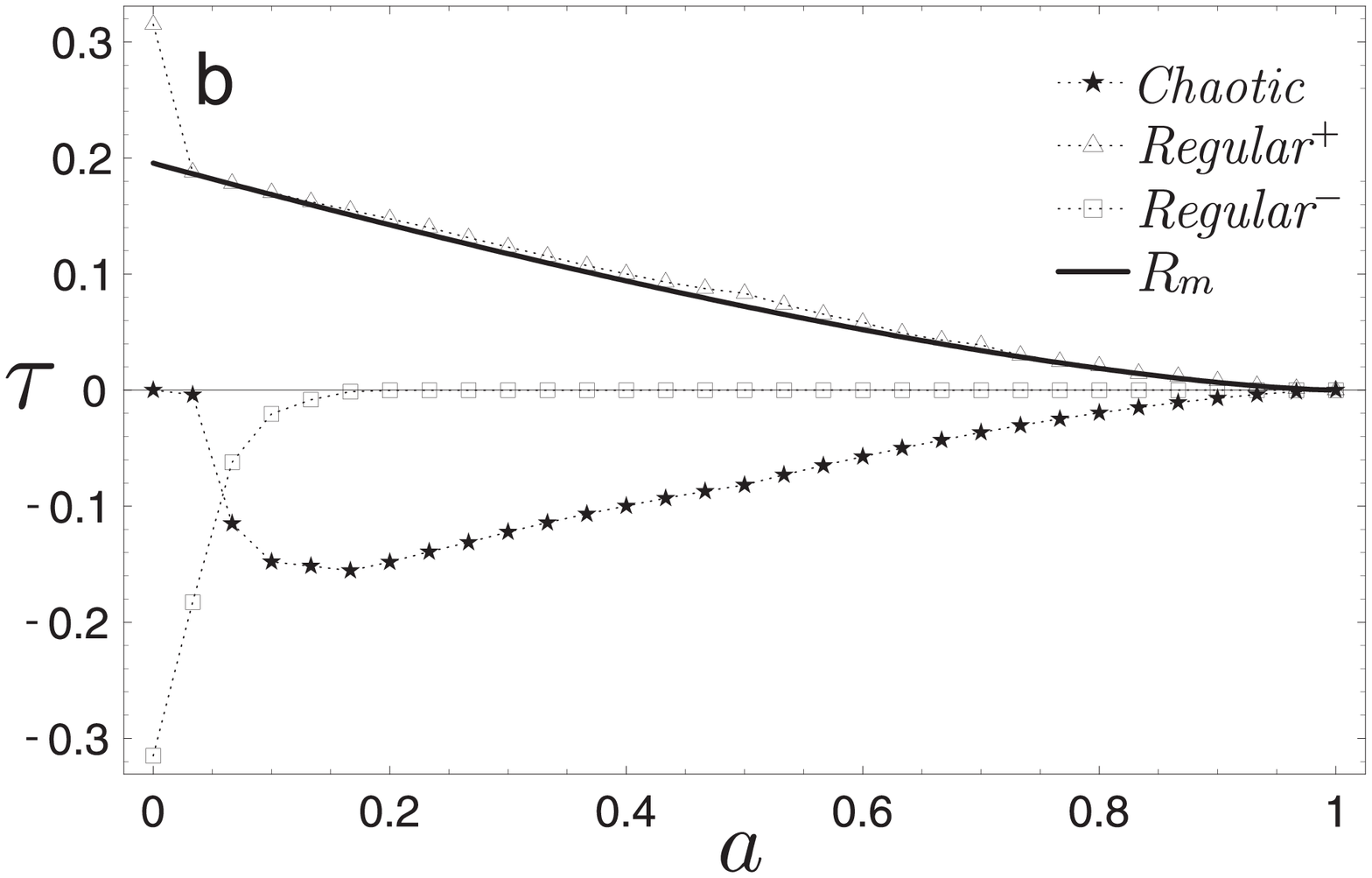}
\caption{Fractions of the energy shell pertaining to different
invariant sets $\M_i$, cf.\ (\protect\ref{eq:EnergyShell}) (a) and
directed current $\tau$ for these invariant sets (b) vs.\ the
barrier size $a$ at $r_{\rm c} = 2$. Solid lines: analytical
solutions, see appendix B.} \label{fi:trans}
\end{figure}

This does not preclude, however, that directed transport occur within
individual invariant sets $\M_i$, such as in particular for the
chaotic sea. In figure \ref{fi:trans}, we compare the
corresponding contributions from the chaotic sea and the regular
components specified above. It is obvious that chaotic transport
reflects directly the imbalance between currents in the regular
regions of phase space. A dominant r\^ole is played by the set of
non-contractible trajectories passing below the side walls, cf.\
figure \ref{fi:po}m. They form a continuous family parameterized,
e.g., by the positions $(x_{\rm c},y_{\rm c})$ of the center of the
circular arcs making up the orbit. Its vertical component, in turn,
is bounded from above by $y_{\rm c} < 1 - a - r_{\rm c}$. This allows
us to estimate analytically the contribution to transport of the
invariant set formed by these orbits, which we denominate $R_m$, as a
function of $a$, see appendix B; the results are included as full
lines in figure \ref{fi:trans}b. For comparison, figure
\ref{fi:trans}a shows  the fractions of the energy shell volume $f_i =
\int_{\M_i}{\rm d}^{2f-1}r \rho_t(\mathbf{r})$ pertaining to each
invariant set $\M_i$ as functions of $a$. We note in passing that
directed transport, say, in the chaotic sea would allow for a finite
current to develop even from an unbiased initial condition (i.e.,
symmetric in $p_x$).

There is no transport in the limit $r_{\rm c} \to \infty$, as it
corresponds to vanishing magnetic field and thus restores
time-reversal invariance. In the opposite limit $r_{\rm c} \to 0$,
the entire phase space is filled with non-transporting circular
orbits. Likewise, for $a = 0$ transport vanishes because the
billiard becomes spatially symmetric, while for $a = 1$, the unit
cell is closed. For $a > 0$, at $r_{\rm c} = 2$, the principal
contribution to transport to the right comes from $R_m \subset$
\emph{Regular$^+$}. The slight discrepancy between the curves
corresponding to \emph{Regular$^+$} and the estimate for $R_m$,
respectively, in figure \ref{fi:trans} reflects the contribution
by the regular island pertaining to the periodic orbit depicted in
figure \ref{fi:po}c. The component \emph{Regular$^0$}, in turn, is
dominated by the periodic orbit in \ref{fi:po}a; its area
decreases from a maximum at $a = 1$ till it vanishes at $a = 0.5$
where the underlying orbit gets pruned. There is no appreciable
regular transport to the left for $a > 0$. The finite values of
$f_{R^-}$ we obtained for small $a$ are numerical artifacts, owed to
the difficulty to distinguish regular trajectories from chaotic ones
with very small positive Lyapunov exponent as they occur in this
range.

The existence of the phase-space component $R_m$ also renders directed
transport very robust with respect to deviations from the symmetry $l
= w$. In either direction, $R_m$ is not affected as it depends only on
$a$. By contrast, components like \emph{Regular$^0$} eventually disappear
with the underlying regular islands for $l \gg w$ or vice versa, but
are replaced by other components with equivalent transport features. Even
in the corresponding limits, the phase-space components supporting
directed transport maintain a finite weight. 

\section{Quantization}
Quantizing a billiard threaded by a magnetic field amounts to solving
the time-independent Schr\"odinger equation including a vector
potential $\mathbf{A}(\mathbf{x})$, $\mathbf{x} = (x,y)$,
\begin{equation}\label{eq:schroedinger}
\frac{1}{2m} \(-i\hbar\nabla - q\mathbf{A}(\mathbf{x})\)^2
\wf(\mathbf{x}) = E\wf(\mathbf{x})
%\frac{1}{2m}\( \(-i\hbar\parcial{}{x}-\frac{eBy}{2} \)^2 +
%\(-i\hbar\parcial{}{y}-\frac{eBx}{2}  \)^2\)\wf(x,y)=(E-V)\wf(x,y)
\end{equation}
with Dirichlet boundary conditions along the walls.

The vector potential $\mathbf{A}(\mathbf{x})$ is determined by the
magnetic field through $\mathbf{B} = \nabla\times\mathbf{A}$ only
up to a gauge. We here choose the Landau gauge \cite{HS02}
\begin{equation}
\mathbf{A}(\mathbf{x})= B\(-y,0\)\label{eq:gaugeLandau}
\end{equation}
as the resulting eigenstates preserve the reflection symmetry
(\ref{eq:reflx}). The resulting Schr\"odinger equation is
equivalent to a Helmholtz equation with the same boundary
conditions, up to a Peierls phase in its interior \cite{KM88}.

Among the quantities to be derived from the wave function we mention
the spatial probability density
\begin{equation}
\rho(\mathbf{x}) = \abs{\wf(\mathbf{x})}^2
\label{eq:probden}
\end{equation}
and the probability density current
%\begin{equation}
%\pc(\mathbf{x}) = -\frac{\hbar}{m}\, {\rm Im}\!\
%(\wf(\r)\nabla_{\mathbf{x}} \wf^*\!(\mathbf{x}) \)-\frac{q}{m}\,
%\vp\!\abs{\wf(\mathbf{x})}^2 \label{eq:probabilityCurrent}
%\end{equation}
\begin{equation}
\mathbf{J}(\mathbf{x})=-\frac{\hbar}{m}\,{\rm
Im}\!\(\wf(\mathbf{x}) \nabla\wf^*\!(\mathbf{x}) \) -
\frac{q}{m}\,\mathbf{A}(\mathbf{x})\!\abs{\wf(\mathbf{x})}^2
\label{eq:probcur}
\end{equation}
which is useful, e.g., to identify vortices, see \ref{se:vort}, and to
compare the quantum-mechanical probability flow with classical
transport mechanisms, see \ref{se:qtrans}.

\subsection{Quantization algorithm}
Numerous algorithms are available for solving
(\ref{eq:schroedinger}) \cite{NT88,WB90,Kla93,WWG94,JB95,BB97,HS00}. 
We choose a finite-element technique \cite{STI89} as it is optimally
suited to  fully exploit the simple geometry of the billiard, with
Dirichlet boundary conditions along straight walls at right angles. A
practical difficulty arises, however, at the free ends of the side
walls that form cusps of the boundary and induce a very non-smooth
behavior of the wave function in their vicinity. In order to achieve
reasonable solutions even in these critical regions, we employ cubic
Hermite polynomials as basis functions in the finite-element algorithm
\cite{Pre75}. The presence of the vector potential is reflected in
a finite Peierls phase corresponding to the magnetic flux through
each cell of the grid underlying the finite-element algorithm
\cite{KM88}.

Another non-standard aspect in this context is the periodicity
(\ref{eq:perix}) of the billiard. It is appropriately taken into
account in the quantum mechanical treatment by working in the Bloch
formalism \cite{AM76}. Assuming a total system size of $M$ unit cells
with overall periodic boundary conditions, it implies in particular a
periodicity condition from cell to cell
\begin{equation}
\wf_{k_x}(x+1,y)=\wf_{k_x}(x,y)\exp(i k_x);
\label{eq:Factordefase}
\end{equation}
where $k_x$ is the Bloch number (equivalent to the quasimomentum and
the Bloch phase) given by
\begin{equation}
k_x=\frac{2\pi m}{L}=\frac{2\pi m}{Ml}
\end{equation}
where $L = Ml$ is the total billiard length and
$m=0,1,2,\ldots,M-1$. This condition allows to solve the
Schr\"odinger equation over an effective area that corresponds to
the unit cell only, by requiring a fixed phase difference given by
(\ref{eq:Factordefase}) between the open boundary sections along
its left and right edges, respectively, where Dirichlet boundary
conditions do not apply.

Resuming, the finite-element approach has to be complemented with (i),
a local phase accumulated around the circumference of each cell of the
grid, and (ii), a global phase difference between the open sections of
the billiard's unit cell. Both phases are to be considered parameters
of the system, corresponding to the magnetic flux and the Bloch
phase, respectively.

Details of the algorithm are deferred to appendix C.

\subsection{Spectrum and eigenstates}

\subsubsection{Landau levels}
\begin{figure}[ht!]
\centering
\includegraphics[width=0.7\textwidth]{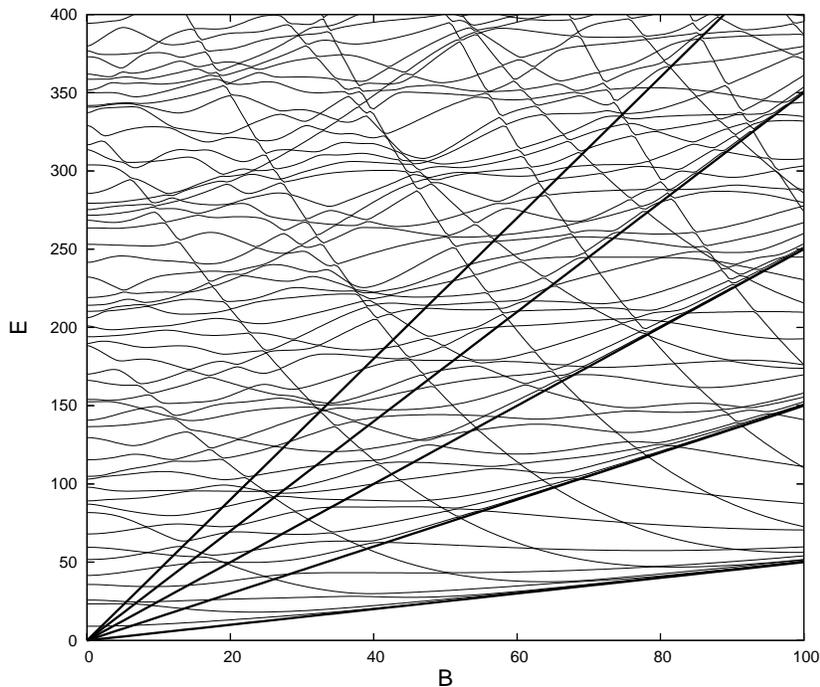}
\caption{Eigenenergies $E$ vs.\ magnetic field $B$ for $a=0.5$, at
$k=0$. Bold straight lines indicate Landau levels
(\protect\ref{eq:landau}).} \label{fi:EvsB}
\end{figure}

For a two-dimensional electron gas, i.e., charged particles moving in
a plane perpendicular to a constant homogeneous magnetic field but
otherwise free, the eigen\-ener\-gies are given by 
\begin{equation}
E_n = (n + 1/2) \hbar qB,
\label{eq:landau}
\end{equation}
known as Landau levels, and the eigenstates are analogous to those
of a two-dimensional symmetric harmonic oscillator \cite{CDL77}.
They are characterized by two spatial scales, one given by the
cyclotron radius $r_{\rm c} = p/qB$, as for the corresponding
classical motion, the other by the magnetic length $b =
\sqrt{2\hbar/qB}$ \cite{Bal98}, the radial width of eigenstates
concentrated along circles of radius $r_{\rm c}$. These Landau states
become the dominant feature of the quantum billiard in the limit of
strong magnetic field, i.e., for $B \gg p/q$ and $B \gg \hbar/q$. This
expectation is confirmed by the data shown in figure \ref{fi:EvsB}.
While for low field, we observe the spectral structures like level
repulsion typical for a classically chaotic system, the eigenenergies
tend to accumulate towards the Landau levels (bold lines) in the
high-field limit. Their influence remains visible in the band
structure as marked concentrations of bands, in particular for strong
field, as visible in figure \ref{fi:bands1}c below.

\subsubsection{Band spectrum}
The fingerprints of directed transport in the band spectrum are
amply discussed in \cite{SO&01,SDK05}. A crucial concept for their
understanding is that of diabatic bands, coherent large-scale
structures in the spectrum that are formed by connecting sections
of actual (``adiabatic'') bands across small gaps (``avoided
crossings''). They can be directly associated to the
current-carrying manifolds in classical phase space. Regular
islands with non-zero winding numbers correspond to approximately
straight lines in the spectrum which, upon connecting them
according to the periodicity of the Brillouin zone, exhibit the
same winding number as the underlying classical island. Diffusion
with superposed drift in the chaotic sea is reflected in wiggly
diabatic bands with an average slope related to the mean transport
velocity in this part of the classical phase space. Moreover,
among the ``chaotic bands'' we observe the phenomenon of level
repulsion characteristic for classically chaotic quantum systems
\cite{Haa00}.

\begin{figure}[ht!]
\includegraphics[height=0.62\textwidth]{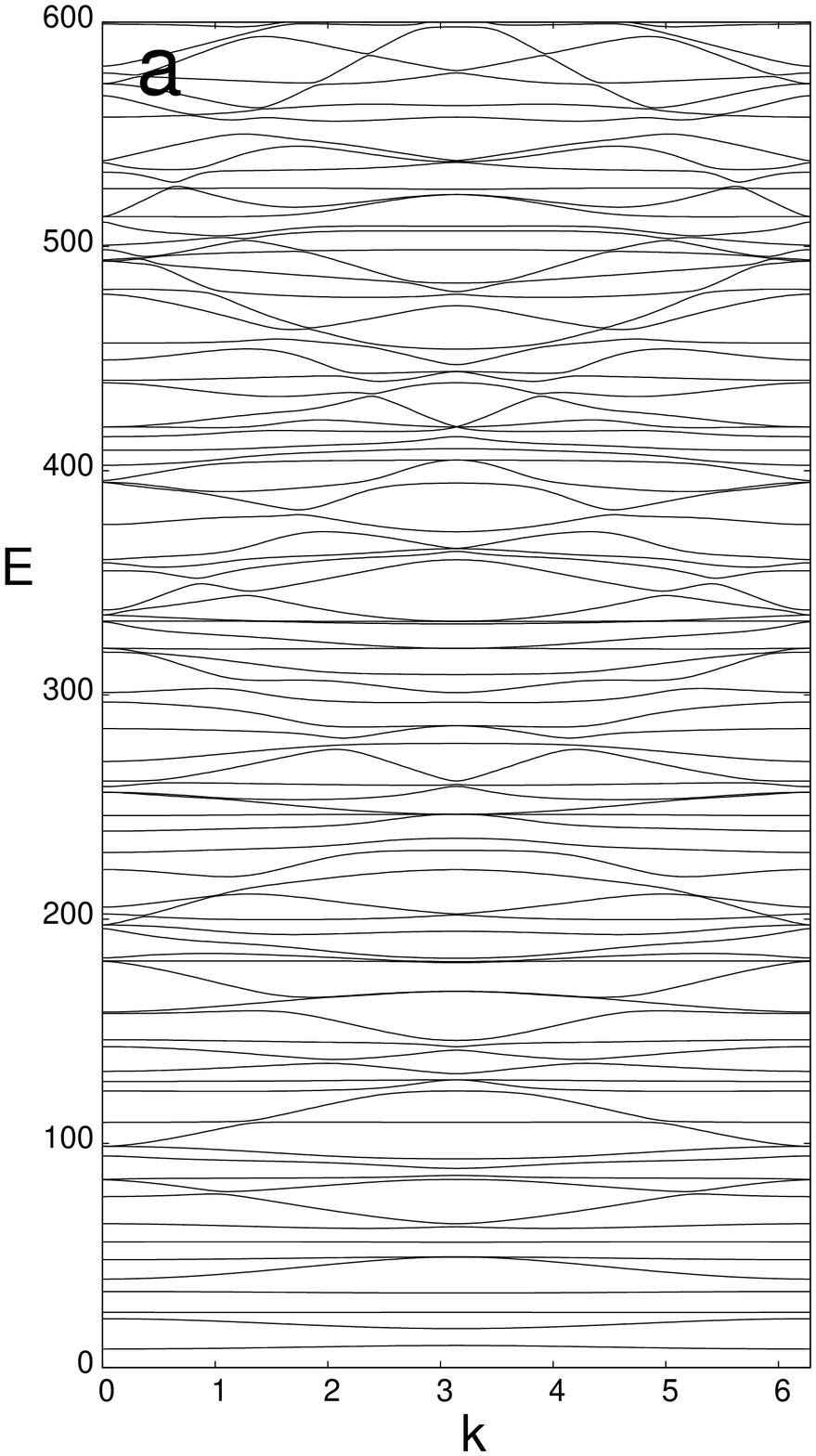}
\includegraphics[height=0.62\textwidth]{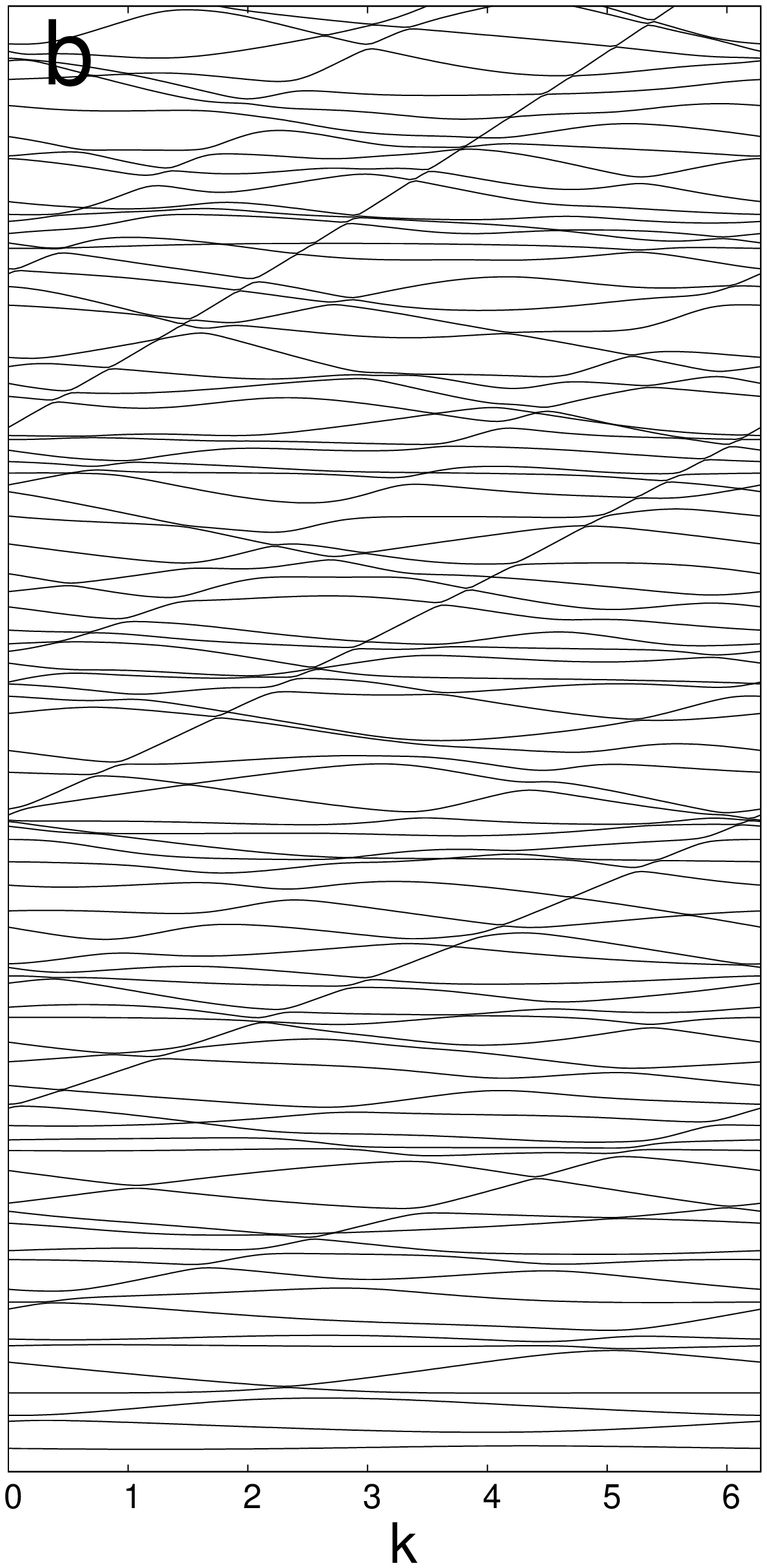}
\includegraphics[height=0.62\textwidth]{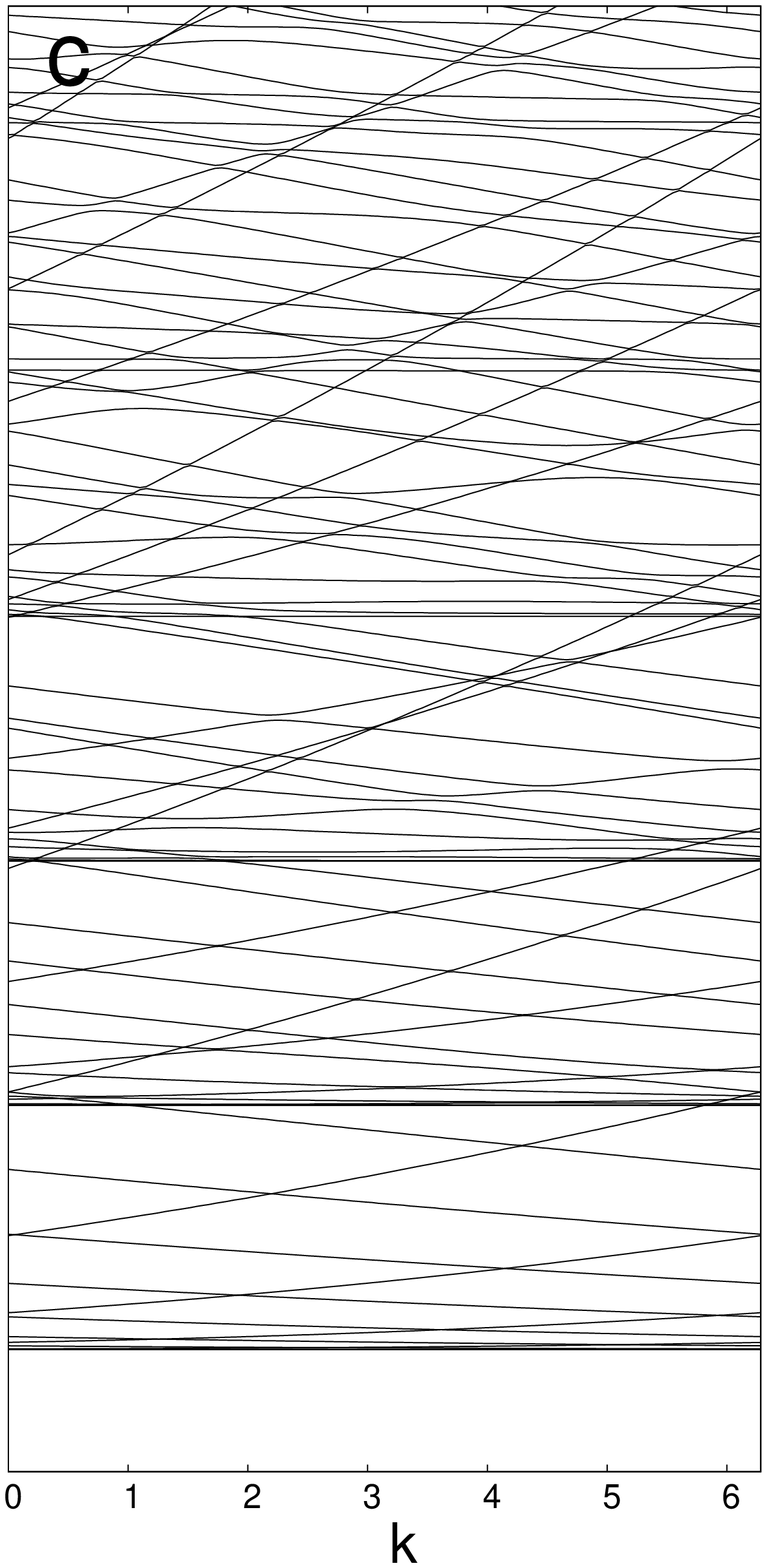}
\caption{Lowest 90 energy bands for $a=0.5$ and $B=0$ (a), $B=10$
(b), and $B=100$ (c).} \label{fi:bands1}
\end{figure}

\begin{figure}[ht!]
\includegraphics[height=0.62\textwidth]{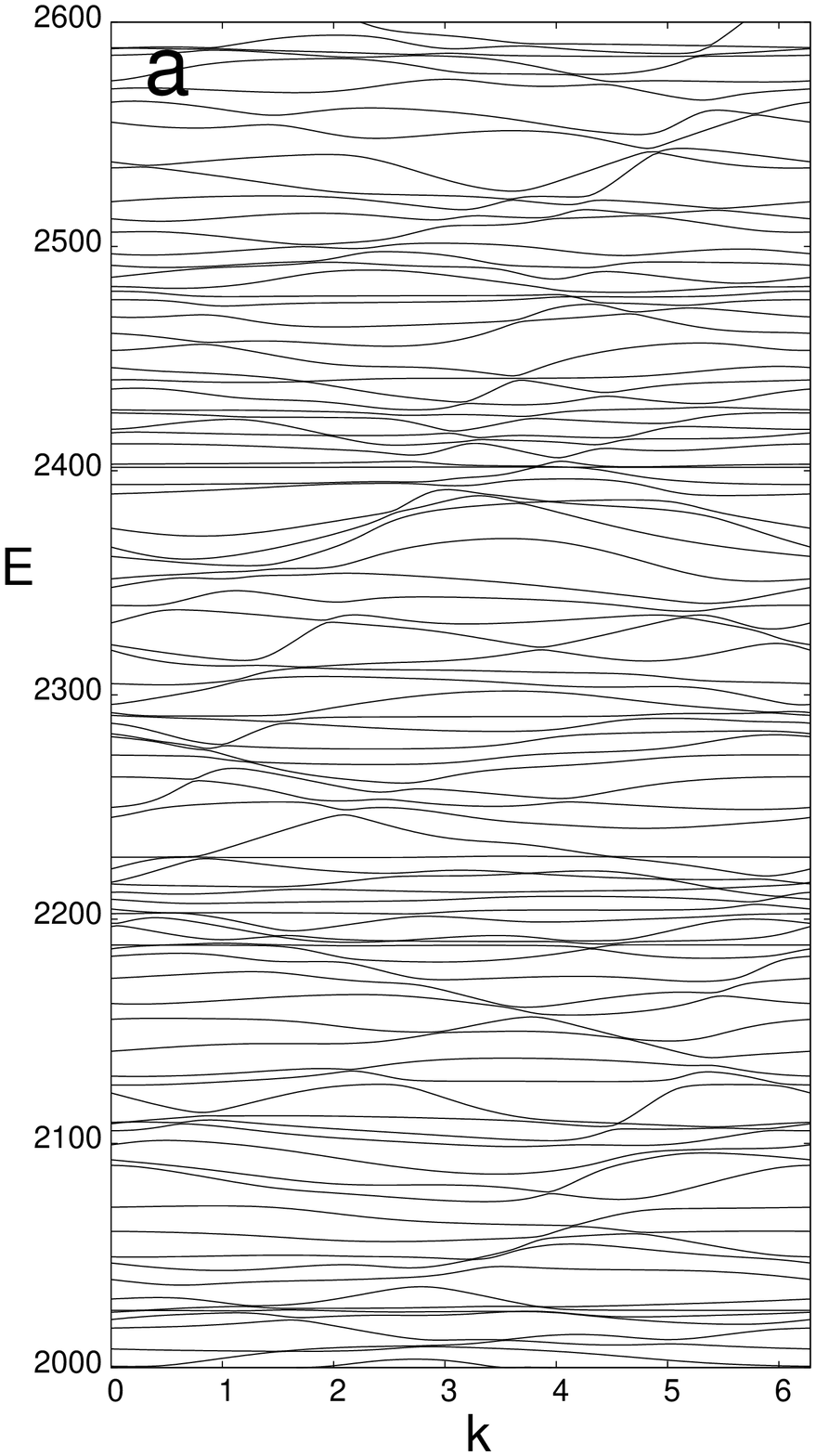}
\includegraphics[height=0.62\textwidth]{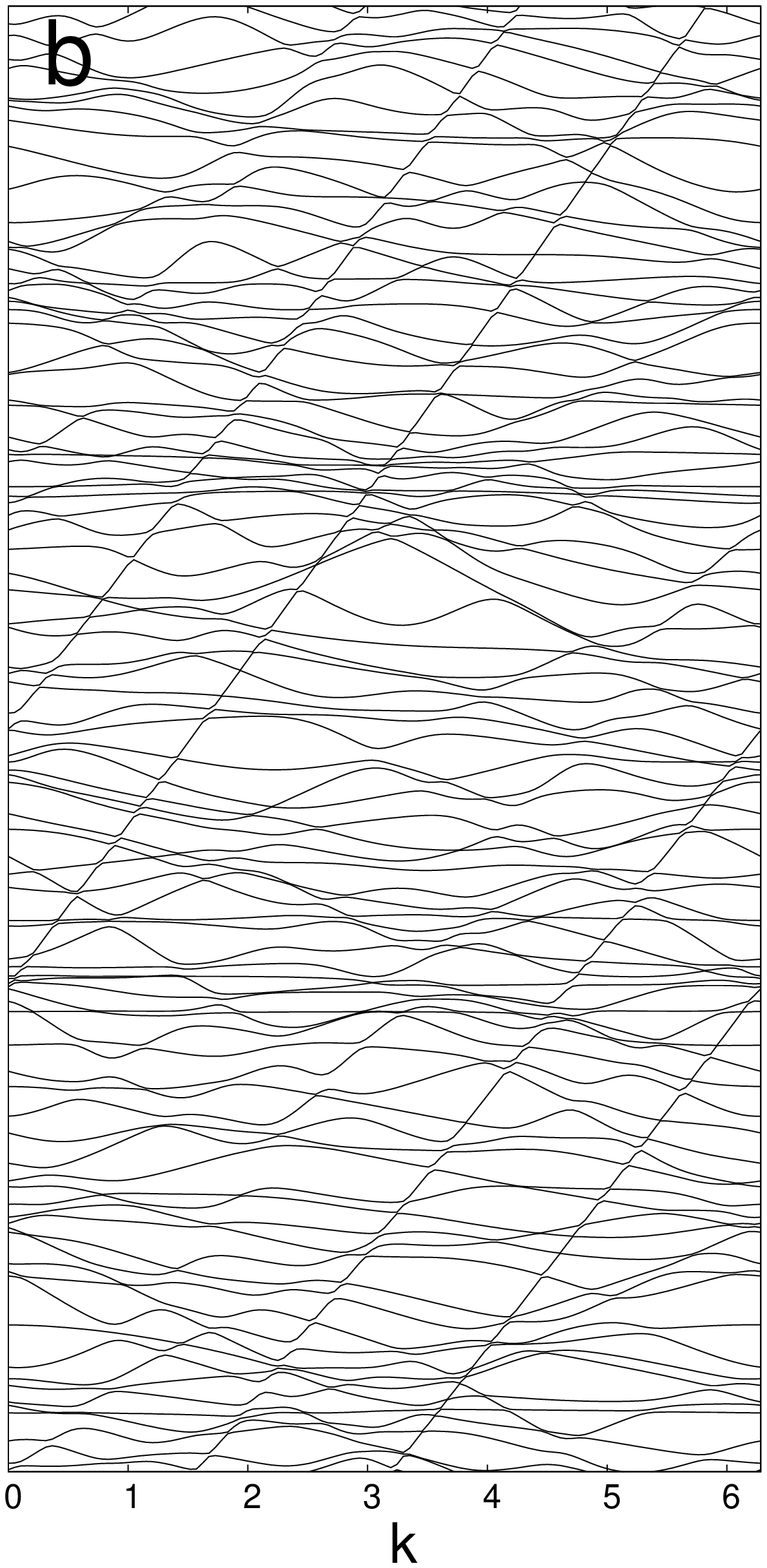}
\includegraphics[height=0.62\textwidth]{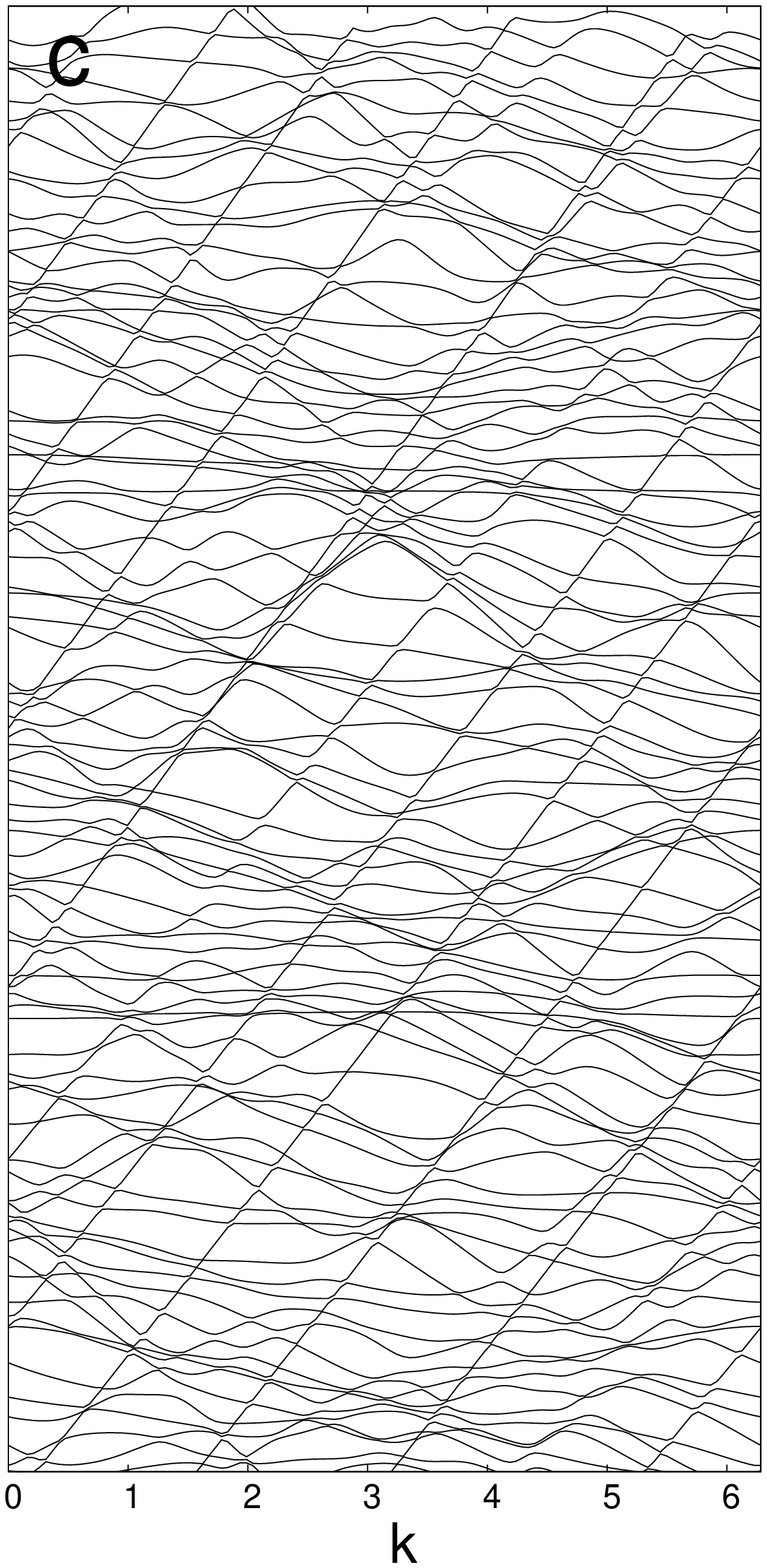}
\caption{Energy bands $E_n$ with $300 \le n \le 400$ for $B = 10$
and $a = 0.8$ (a), $0.5$ (b), and $0.2$ (c).} \label{fi:bands2}
\end{figure}

As mentioned in \ref{se:dirtrans}, a dominating feature of
directed transport in the magnetic billiard is the contribution by
non-contractible open trajectories that correspond to quasi-free
motion through the gaps left between the side walls and the lower
wall of the channel. As discussed above, the winding numbers of
these trajectories form a con\-tinu\-um and depend on energy, by
contrast to the regular islands whose rational winding numbers reflect
their topology and therefore may be called structurally stable.
Indeed, we ob\-ser\-ve structures in the band spectrum, cf.\ figures
\ref{fi:bands1}b-c, that resemble the parabolic bands
pertaining to quasi-free particles moving in a crystalline
potential \cite{AM76}. The conspicuous, apparently straight lines
with positive slope appearing in figures \ref{fi:bands2}b-c are
closer to features seen in quantum ratchets \cite{SO&01,SDK05} but
in fact also show a slight curvature. They belong to the same
category as those in figures \ref{fi:bands1}b-c and can be
associated to the classical phase space component $R_m$. At the
same time, we do observe some approximately straight diabatic
bands with negative slope that pertain to very small regular
islands with winding number $< 0$. By contrast, a bias of the mean
slope of the remaining wiggly bands is not so easy to discern with
the unaided eye. These features have in common that they are
incompatible with the familiar reflection symmetry of energy bands
with respect to $k = \pi$ (as in figure \ref{fi:bands1}a)---this
is indeed how the breaking of TRI by the magnetic field is
reflected in the band structure. They are of particular relevance
for directed transport, as discussed in section \ref{se:qtrans}
below.

\subsubsection{Bloch states}
\begin{figure}[ht!]
\includegraphics[width=\textwidth]{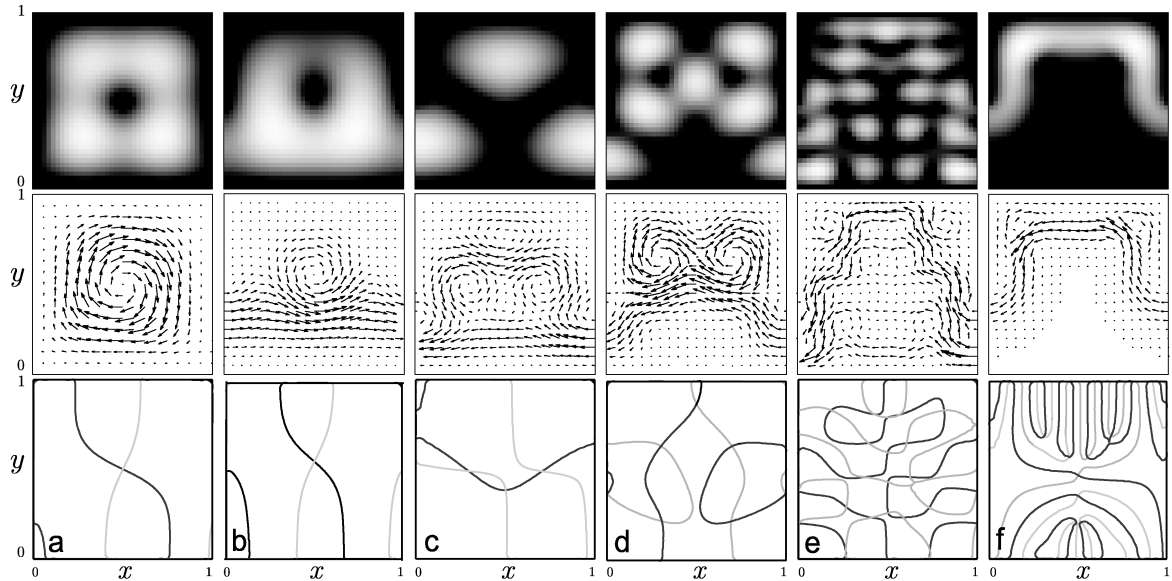}
\caption{Examples of eigenstates represented by their probability
density (\protect\ref{eq:probden}) (upper row, greyscale from
black $\equiv$ zero to white $\equiv$ maximum) and flow,
(\protect\ref{eq:probcur}) (middle row), and nodelines (lower row)
of the real (black) and imaginary parts (grey) of the
wavefunction. Parameters are, $B=10$, (a-e), $100$ (f), $a = 0.8$
(a), $0.5$ (b-f), $k = 0$ (a-c), $\pi$ (d-f), and energy level
index, from a to f, $n = 3$, $2$, $3$, $8$, $24$, $11$.}
\label{fi:states1}
\end{figure}

\begin{figure}[ht!]
\includegraphics[width=\textwidth]{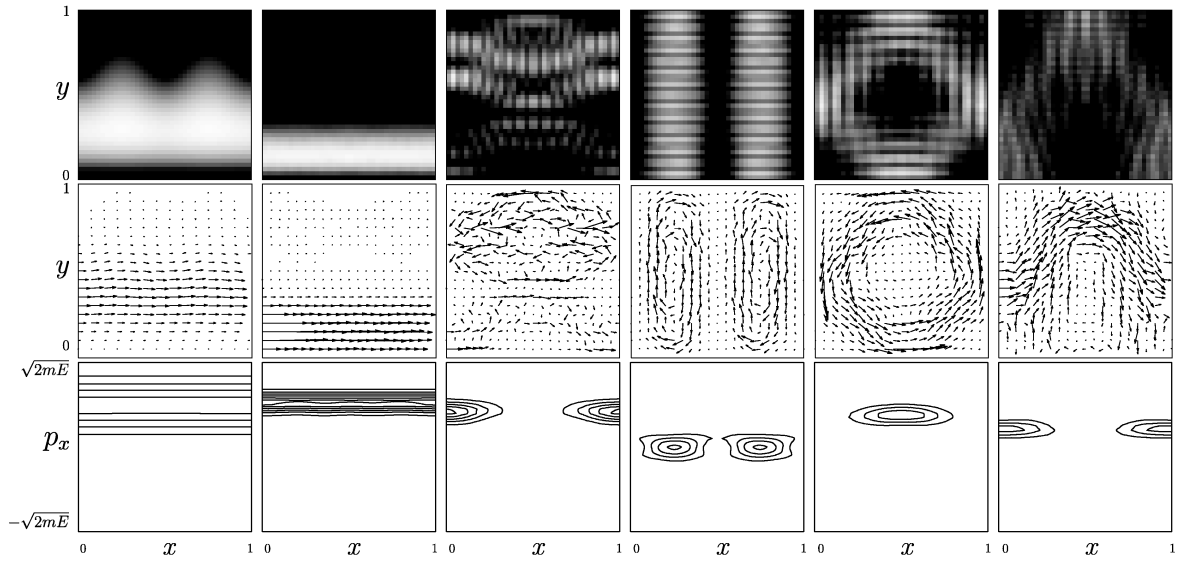}
\caption{Examples of eigenstates represented by their probability
density (\protect\ref{eq:probden}) (upper row, greyscale from
black $\equiv$ zero to white $\equiv$ maximum) and flow,
(\protect\ref{eq:probcur}) (middle row) and Husimi functions in
the plane $(x,p_x)$ along the lower wall of the billiard (lower
row), see text for details. Parameters are $B=10$, $k=0$ and, from
a to f, $a = 0.25$, $0.5$, $0.5$, $0.5$, $0.8$, $0.2$, and energy
level index $n = 2$, $128$, $167$, $169$, $201$, $322$,
corresponding to eigenenergies
% $E = 20.25$, $860.58$, $1126.64$, $1134.61$, $1349.54$, $2101.20$.
$E = 20.25$, $860.6$, $1126$, $1134$, $1349$, $2101$.
}\label{fi:states2}
\end{figure}

We analyze the eigenstates of the periodic billiard, that is, its
Bloch states, to resolve the flow structures underlying global
transport properties, to identify fingerprints of the corresponding
classical dynamics, and to observe vortices of the quantum flow as
special quantum features associated with directed transport in the
billiard \cite{SK&99}. In figures \ref{fi:states1} and
\ref{fi:states2}, we simultaneously show the probability density and
flow as well as node lines of the real and imaginary parts of the wave
function and its phase-space representation in terms of Husimi
functions.

The Husimi functions presented in the third row of figure
\ref{fi:states2} have been constructed following the approach
detailed in \cite{CR93,LN&96}. It reduces the system's
four-dimensional phase space, exploiting the approximate
factorization of the wave function near the lower horizontal wall
(sector I in figure \ref{fi:shape}) into $x$- and $y$-dependences,
to a Poincar\'e surface of section spanned by the two dimensions
$x$ and $p_x$.

The density distributions already allow us to classify the states
according to a number of pertinent categories. For parameter values
close to the corresponding limits, we observe states that resemble the
solutions of the Schr\"odinger equation for a closed square
billiard (\ref{fi:states1}a) or a free channel
(\ref{fi:states2}b), respectively. In particular, so-called
bouncing-ball states (\ref{fi:states2}d) pertain to these
categories. Even for relatively large openings between the cells,
we find states with a marked minimum across the gap
(\ref{fi:states1}b) which therefore belong to the first category
mentioned above. By contrast, the continuity equation for the
probability flow implies that states supporting strong directed
transport tend to exhibit high probability density and strong flow
through the openings (\ref{fi:states1}c,d, \ref{fi:states2}a,b). Where
the flow remains well bundled even within the cell, it tends either to
wind along the upper wall and sidewalls (\ref{fi:states1}f), if
transport is to the left, or to adhere to the lower wall
(\ref{fi:states2}b), in the opposite case. We can characterize these
states as edge states \cite{HS02,Hal82,Kra98}, borrowing a concept
from the context of the quantum Hall effect. In classical terms, in
turn, they can be associated to so-called skipping orbits that creep
along the walls, as they occur in the high-field regime, cf.\ figure
\ref{fi:po}l.

Energy eigenstates with probability densities and flows concentrated
near the lower wall (\ref{fi:states2}a,b) are related to the
phase-space component $R_m$. Indeed, their current patterns resemble
the corresponding trajectories (figure \ref{fi:po}m) and their Husimi
functions tend to localize over the corresponding invariant
tori. Other fingerprints of the classical phase-space flow can be
observed in the Husimi functions in figures
\ref{fi:states2}c,d,e,f as they concentrate over the orbits in figures
\ref{fi:po}c,g,a,i, respectively.

\subsubsection{Vortices, dia- and paramagnetism}\label{se:vort}
By analogy to a hydrodynamic flow with the same boundary
conditions, one expects vortices to occur in the
quantum-mechanical probability flow in the range $0 < a < 1$, in
particular for $a \applss 1$. Experimental results for microwave
billiards as systems analogous to quantum billiards
\cite{SK&99,KH&07} and theoretical results for mesoscopic systems
\cite{ES&98} support this expectation. Writing the quantum
probability current (\ref{eq:probcur}) in the form
\begin{equation}
\mathbf{J}(\mathbf{x}) = \rho(\mathbf{x})
\(\hbar\nabla S(\mathbf{x}) - q\mathbf{A}(\mathbf{x})\),
\label{eq:probcurfac}
\end{equation}
where $S(\mathbf{x})$ denotes the phase of the wavefunction,
we identify the second factor on the rhs as a velocity
\begin{equation}
\mathbf{v}(\mathbf{x}) = \hbar\nabla S(\mathbf{x}) -
q\mathbf{A}(\mathbf{x}).
\label{eq:probvel}
\end{equation}
This expression shows that the magnetic field contributes to the
formation of vortices directly and independently of the quantum phase.
The difference between the two contributions is brought out even more
clearly if we consider the corresponding vorticity (synonymous to the
fluxoide in the context of superconductivity), defined as the
line integral of $\mathbf{v}(\mathbf{x})$ around a closed contour
$\Gamma$,
\begin{equation}
\oint\limits_{\Gamma}{\rm d}\mathbf{x}\cdot\mathbf{v}(\mathbf{x}) =
\hbar\oint\limits_{\Gamma}{\rm d}\mathbf{x}
\cdot\nabla S(\mathbf{x}) - q\oint\limits_{\Gamma}{\rm d}\mathbf{x}
\cdot\mathbf{A}(\mathbf{x}) = q\(\Phi_l - \Phi_{\rm mag}\).
\label{eq:probveloint}
\end{equation}
Here, $\Phi_{\rm mag}$ is the magnetic flux of the external field
enclosed in the contour. By contrast, the term $\Phi_l$ indicates
topological defects of the phase of strength $2\pi l$, $l$
integer. This discretization is equivalent to the quantization of
the magnetic flux $\Phi_l = 2l\pi\hbar/q$ generated by the vortex.
This induces a different behavior of the vorticity for the two
types of vortices upon contracting the contour towards the center
of the vortex: In the absence of singularities of the external
field, $\Phi_{\rm mag}$ then reduces to zero and so does the
vorticity, while at topological defects the vorticity remains finite
in the limit and therefore requires $\mathbf{v}$ to diverge. In order
that the flow $\mathbf{J} = \rho\mathbf{v}$ remain finite in turn,
this implies that the density $\rho(\mathbf{x})$ decay with
decreasing distance $r$ from the vortex at least as $r$, that is,
possess a first-order zero at the center, providing another
criterion to distinguish the two classes.

Moreover, topological defects can occur with either sign,
anti-clockwise or clockwise, corresponding to a paramagnetic or
diamagnetic response to the external field, respectively. The sign of
$\Phi_{\rm mag}$ depends on the orientation of the magnetic field; it
always gives rise to diamagnetic response, that is, in our case, to
clockwise vortices.

\begin{figure}[ht!]
\includegraphics[width=\textwidth]{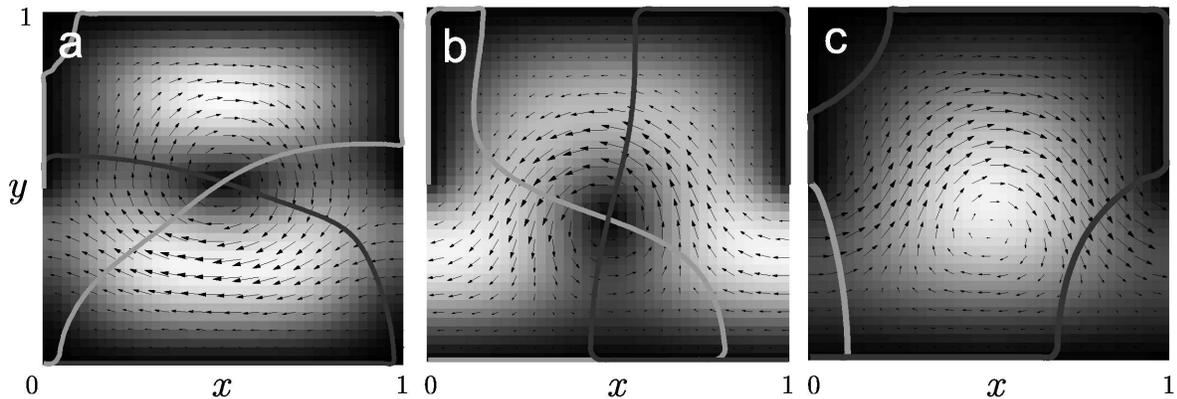}
\caption{Examples of eigenstates featuring a diamagnetic (a) and a
paramagnetic (b) topological defect, and a diamagnetic
non-quantized vortex (c), represented by their probability density
(\protect\ref{eq:probden}) (greyscale from black $\equiv$ zero to
white $\equiv$ maximum) and flow (\protect\ref{eq:probcur}),
superposed with nodelines of the real (black) and imaginary parts
(grey) of the wavefunction. Note the coincidence of the quantized
vortices (panels a,b with quantum numbers $l = -1$, $1$, resp.)
with zeros of the density and nodeline crossings (phase defects).
Parameters are $B=10$, $a=0.5$, $k=\pi$ and, from a to c, energy
level index $n = 3$, $2$, $1$.} \label{fi:vort}
\end{figure}

In figure \ref{fi:vort}, we show three eigenstates featuring the three
classes of vortices mentioned above, a diamagnetic (panel a) and a
paramagnetic (b) topological defect, and a non-quantized diamagnetic
vortex (panel c). We observe that indeed, the topological defects
are associated with isolated nodes of the density, while the
non-quantized vortex typically sits close to a maximum of
$\rho(\mathbf{x})$. Superposing graphs of the flow
$\mathbf{J}(\mathbf{x})$ with nodelines of ${\rm Re} \psi(\mathbf{x})$
(black bold lines) and ${\rm Im} \psi(\mathbf{x})$ (grey)
demonstrates that nodeline crossings, i.e., phase defects, coincide
with vortices. Except for their intersections, these nodelines are
gauge dependent and therefore have no physical significance. Their
orientation and density, however, reflects the probability flow and
serves as an alternative way to visualize it. For an energy eigenstate
featuring both types of vortices, see figure \ref{fi:states1}c. In our
case, and by contrast, e.g., to microwave billiards \cite{KH&07},
there is no reason for vortices to occur pairwise (one dia-, one
paramagnetic) or to be accompanied by a pair of saddles (hyperbolic
points) of the flow. Our system therefore corresponds to a different
universality class as concerns vortex statistics. We shall however not
pursue this subject any further.

\subsection{Directed quantum transport}\label{se:qtrans}
On general grounds based on the correspondence principle, we expect to
find evidence of directed transport in the quantized billiard, of
comparable magnitude as in its classical counterpart, at least in the
semiclassical parameter regime. This expectation has been amply
confirmed in the case of ratchets \cite{SO&01,SDK05}. Moreover, the
examples of probability densities and probability flows in figures
\ref{fi:states1} and \ref{fi:states2} clearly show that features of
the classical dynamics relevant for transport, e.g., periodic orbits,
are reflected in the quantum system. In particular, the phase-space
component $R_m$ decisive for the occurrence of classical directed
currents also dominates quantum transport and renders it robust, e.g.,
against deviations from square unit cells $l = w$.

\begin{figure}
\centering
\includegraphics[width=0.6\textwidth]{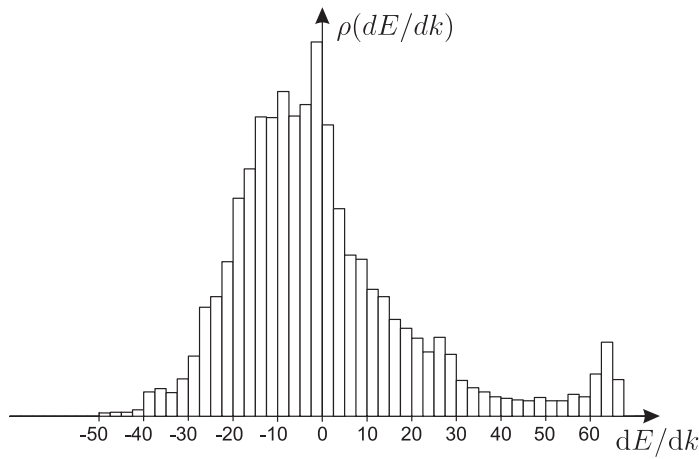}
\caption{Histogram of the band slope (\protect\ref{eq:bandslope}) for
energy bands $E_n$ with $300 \le n \le 400$ at $B = 10$ and $a = 0.2$,
as shown in figure \protect\ref{fi:bands2}c.}
\label{fi:bandstat}
\end{figure}

Adopting an approach developed in \cite{SO&01,SDK05}, we here present
less direct though more comprehensive evidence for directed
transport. According to Bloch theory, the mean velocity of a
wavepacket prepared in the energy band $E_n(k)$ at quasimomentum $k$
is given by the band slope at this point \cite{AM76},
\begin{equation}
\bar{v}_n(k) = \frac{1}{\hbar} \frac{{\rm d}E_n(k)}{{\rm d}k}.
\label{eq:bandslope}
\end{equation}
Directed transport can therefore be assessed on basis of the band
structure, in particular of the statistics of band slopes (level
velocities). In figure \ref{fi:bandstat} we present a histogram of
band slopes obtained for the same parameter values and the same energy
window as underlies figure \ref{fi:bands2}c. The two most prominent
features of the band structure are clearly reflected in the histogram,
approximately straight diabatic bands with a positive slope $\approx
66$, corresponding to the phase-space component $R_m$ of the classical
system, and wiggly bands with negative slope on average, corresponding
to chaotic drift towards the left: They appear as a relatively marked
peak at the expected value $\bar{v} \approx 66$, in very good
agreement with the mean velocity $\bar{v}_m = 66.16$ of the component
$R_m$ in the relevant energy range $2000 \leq E \leq 2600$, and
a broad approximately Gaussian background biassed towards $\bar{v} <
0$, respectively. The global mean velocity vanishes within numerical
accuracy, reflecting the quantum analogue of the classical sum rule
(\ref{eq:sumrule}) for directed transport \cite{SO&01,SDK05}. The
conspicuous peak at $\bar{v} = 0$ may be associated to the
contribution of the avoided crossings. All in all, our histogram
closely resembles the analogous figure 7 in \cite{SDK05} (except for a
sign reversal of the velocities).

\section{Conclusion}
Billiards emerged as mathematically-minded models for classical and
quantum complex dynamics. Yet they inspired a surprisingly versatile
range of laboratory experiments, ranging from lateral semiconductor
superlattices \cite{WR&91} to microwave billiards \cite{SS90,GH&92}.
It is therefore not surprising that a more recent development
in classical complex dynamics, directed transport in Hamiltonian
ratchets, is also making its way from one-dimensional but
time-dependent models \cite{SO&01,SDK05} to billiards, with their
specific mathematical beauty and diverse experimental applications.

In this case, two of the functions fulfilled in ratchets by the
external driving are taken over by a static magnetic field, namely
rendering the dynamics partially chaotic and breaking time-reversal
invariance. This allows to keep the geometry of the billiard
exceedingly simple, a setup readily sketched on squared paper. The
elementary shape also greatly facilitates the quantization of the
system, using finite-element methods that appear to be taylor-made for
this application.

The mechanisms responsible for directed transport in the classical
billiard proposed in this paper largely carry over to the quantum
system and are manifest, for example, in its band structure. A
consequence of quantum transport not immediately anticipated from the
classical dynamics is, e.g., the occurrence of vortices of the
probability flow that coincide with phase defects of the wavefunction
and correspond to a dia- or paramagnetic response of the system,
respectively.

This work leaves ample space for complementary research. By adding an
ac driving, we can generate a hybrid between an autonomous magnetic
billiard and a ratchet. Even an adiabatic modulation of the magnetic
field strength, without invoking any electric field, would open
further possibilities to attain directed transport.

The restriction of the infinite billiard chain to a section of finite
length with leads attached to both sides yields a scattering system, a
configuration closer to experimental realizable setups, for example a
ballistic two-dimensional electron gas in a semiconductor superlattice
imposed by gate voltages or by lithography. As a closely related
aspect, this setup would provide an additional way to break TRI, by
incoherent decay processes occurring in the electron reservoirs
coupled to the system, and therefore work even in the absence of the
magnetic field.

In such a setup, it would even make sense to study aspects of
trans\-port under an external dc voltage bias, indicating an
alternative way towards the ex\-pe\-ri\-men\-tal observation of
fingerprints of directed transport as it occurs without driving: The
response to a dc voltage would make contact with magnetotransport and
mag\-ne\-to\-re\-sis\-tance.

\ack We enjoyed fruitful discussions with N Berglund, B
Dietz-Pilatus, F Leyvraz, A Richter, H Schanz, T Seligman, and C
Viviescas. Financial support by Volkswagen Foundation (grant
I/78235) and Colciencias (grant 1101-05-17608) is gratefully
acknowledged. One of us (TD) thanks for the warm hospitality and
inspiring atmosphere enjoyed at Centro Internacional de Ciencias
AC in Cuernavaca (Mexico) and at Institut f\"ur Kernphysik,
Technische Universit\"at Darmstadt (Germany).

\appendix
\setcounter{section}{1}

\section*{Appendix A: Classical Poincar\'e map}
\label{ap:map}
To simulate the dynamics of the billard, we employ a geometric
algorithm based on the coordinates $(\mathbf{x},\alpha)$,
constructed so as to minimize rounding errors. In order to obtain the
Poincar\'e map, we need in addition the two transformations that lead
from $(\mathbf{x},\alpha)$ to the Birkhoff coordinates
$(s,\cos\alpha)$ and v.v. The algorithm consists of the following
principal steps:
\begin{itemize}
\item For a given initial condition $(\mathbf{x}_0,\alpha_0)$ and a
fixed value of the cyclotron radius $r_{\rm c}$, calculate the
coordinates $\mathbf{x}_{\rm c}$ of the circle on which the particle
is moving.
\item Find the set $\{\mathbf{x}_l\}$ of intersections of this circle
with the circumference of the unit cell and its periodic continuation,
i.e., with horizontal walls (lines $y=0$ and $y=1$), vertical walls
($x=n$, $n$ integer, with $1-a \leq y < 1$) and the open unit-cell
boundary ($x=n$ with $0 \leq y < 1-a$).
\item Identify among the $\mathbf{x}_l$ the value, call it
$\mathbf{x}_{l_0}$, that corresponds to the initial condition
$\mathbf{x}_0$ itself, taking into account that due to rounding
errors, $\mathbf{x}_{l_0}$ and $\mathbf{x}_0$ might be not exactly
identical.
\item Calculate the angles $\beta_l$ between the vector
$\overrightarrow{\mathbf{x}_{\rm c} \mathbf{x}_0}$ and each one of the
vectors $\overrightarrow{\mathbf{x}_{\rm c} \mathbf{x}_l}$ with $l\neq
l_0$, and rearrange the intersections in the order of increasing
$\beta_l$, from $\beta_{l_0}$ onwards.
% (see figure \ref{fig:algoritmoMEP}).
\item Select the first intersection in forward direction along the
trajectory, i.e., the one pertaining to $\beta_1$ (the smallest after
reordering), as the actual next inflection point of the trajectory. If
it corresponds to a reflection at a wall, the angle at exit is
$\alpha_1 = 2\psi_1 + \beta_1 - \alpha_0$ where $\psi_1$ defines the
orientation of the wall hit by the trajectory. If it corresponds to
passing to the left or right adjacent cell, the angle at exit is
$\alpha_1=\alpha_0 - \beta_1$.
\end{itemize}
Note that with growing cyclotron radius, the number of intersections
to be taken into account in the second step increases, rendering the
algorithm slower.

\section*{Appendix B: Component $R_m$}
\label{ap:CompRm} The phase-space component $R_m$ is formed by
skipping orbits that only hit the lower wall, cf.\ figure
\ref{fi:po}m. For a given value of $r_{\rm c}$, this family is
bounded by those trajectories that touch the lower tip of the
vertical walls at their climax. Taking this condition into
account, one readily shows that $R_m$ is delimited within the
energy shell by
\begin{equation}
R_m=\left\{\;\mathbf{r}\;|\quad y<a', \quad|\alpha|<\arccos
\big(1-(a'-y)/r_{\rm c}\big)\right\},
\label{eq:R1definition}
\end{equation}
where $\mathbf{r}=\{x,y,\alpha\}$ is a point on the energy shell
and $a'=1-a$. Starting from (\ref{eq:R1definition}), we can
estimate analytically the fraction of the energy shell occupied by
$R_m$ as
\begin{equation}
f_{R_m} = (a'-r_{\rm c})\left(\frac{1}{2} + \frac{1}{\pi}
\arcsin\left(\frac{a'-r_{\rm c}}{r_{\rm c}}\right)\right) +
\frac{1}{\pi}\sqrt{a'(2r_{\rm c}-a')}.
\end{equation}
Evaluating (\ref{eq:transEnergyShell}) for the component $R_m$, we
obtain for its contribution to directed transport
\begin{equation}
\fl\qquad\quad\tau_{R_m} = \frac{\sqrt{2E}}{2\pi}
\sqrt{\frac{2a'}{r_{\rm c}} - \frac{a'^2}{r_{\rm c}^2}}
\left(a'-r_{\rm c}-\frac{2r_{\rm c}^2}{\sqrt{a'(2r_{\rm c}-a')}}
\arctan\left(-\sqrt{\frac{a'}{2r_{\rm c}-a'}}\right)\right).
%\tau_{\Rm}=\lim_{T\to\infty}\frac{1}{T}\int_0^T dt
%\int_{\M}d\X\;\hat{F}^t\,\rho_0(\X)\;\chi_{\Rm}\;v_x.
\end{equation}

\section*{Appendix C: Finite-element quantization algorithm}
\label{ap:quant} In order to solve the time-independent
Schr\"odinger equation for the billiard, we cover the unit cell
with a homogeneous square grid of $N\times N$ nodes and express
the wavefunction in terms of bicubic Hermite polynomials
\cite{Pre75}, centered at each node. Substituting this expression
in (\ref{eq:schroedinger}) and applying Galerkin's criterion
\cite{Pre75} leads to a generalized eigenvalue problem where the
Hamiltonian and overlap matrices have dimensions of the order of
$4N^2$. We solve it numerically using the LAPACK routine ZHEGVX.
The periodic boundary conditions with Bloch phase are taken into
account as additional phase factors at those nodes that coincide
with the open sections II and VI of the cell boundary, see figure
\ref{fi:shape}.

This method is subject to limitations both in energy and in magnetic
field strength. The former arise by the small-scale oscillations of
high-lying eigenstates which eventually are no longer resolved by the
mesh. The latter comes about by the typical extension of Landau states,
inversely proportional to $B$, which also leads to a trade-off with the
mesh size.

\end{document}